\documentclass[10pt,twocolumn,aps,prb,showpacs,superscriptaddress,citeautoscript]{revtex4-1}
\usepackage{graphicx}  
\usepackage{dcolumn}   
\usepackage{bm}        
\usepackage{amssymb}  
\usepackage{amsmath}

\newcommand{\beq}{\begin{equation}}
\newcommand{\eeq}{\end{equation}}
\newcommand{\fr}{\bm{r}}
\newcommand{\fG}{\bm{G}}
\newcommand{\fu}{\bm{u}}
\newcommand{\fv}{\bm{v}}

\begin{document}
\title{Resonant and Non-Local Properties of Phononic Metasolids}
\author{Daniel Torrent}
\email{torrent@crpp-bordeaux.cnrs.fr}
\affiliation{Centre de Recherche Paul Pascal, UPR CNRS 8641, Universit\'e de Bordeaux, 115 Avenue Schweitzer, 33600 Pessac, France}
\author{Yan Pennec}
\author{Bahram Djafari-Rouhani}
\affiliation{Institut d'Electronique, de Micro\'electronique et de Nanotechnologie, UMR CNRS 8520, Universit\'e de Lille 
1, 59655 Villeneuve d'Ascq, France}

\date{\today}

\begin{abstract}
We derive a general theory of effective properties in metasolids based on phononic crystals with low frequency resonances. We demonstrate that in general these structures need to be described by means of a frequency-dependent and non-local anisotropic mass density, stiffness tensor and a third-rank coupling tensor, which shows that they behave like a non-local €œWillis medium. The effect of non-locality and coupling tensor manifest themselves for some particular resonances whereas they become negligible for other resonances. Considering the example of a two-dimensional phononic crystal, consisting of triangular arrangements of cylindrical shells in an elastic matrix, we show that its mass density tensor is strongly resonant and anisotropic presenting both positive and negative divergent values, while becoming scalar in the quasi-static limit. Moreover, it is found that the negative value of transverse component of the mass density is induced by a dipolar resonance, while that of the vertical component is induced by a monopolar one. 
Finally, the dispersion relation obtained by the effective parameters of the crystal is compared with the band structure, showing a good agreement for the low-wave number region, although the non-local effects are important given the existence of some resonant values of the wave number.
\end{abstract}
\maketitle
\section{Introduction}
Metamaterials are artificial structures with unusual constitutive parameters not found in natural materials\cite{zheludev2011roadmap}, like negative  compressibility\cite{fang2006ultrasonic}, refractive index\cite{shelby2001experimental,smith2004metamaterials,brunet2015soft} or anisotropic mass density\cite{torrent2008anisotropic}. These properties offer new insights for the propagation of classical waves, and a wide variety of effects and applications have been found, like cloaking shells\cite{schurig2006metamaterial,cummer2007one}, super-lenses\cite{pendry2000negative}, optical and acoustical black holes\cite{narimanov2009optical,cheng2010omnidirectional,Climente2012} or gradient index lenses\cite{GRINLin,lenteTTWu}.

Metamaterials for acoustic or elastic waves, also named metafluids or metasolids, respectively, have been mainly implemented by means of sonic and phononic crystals, which consist of periodic arrangement of inclusions in a fluid (sonic crystal) or elastic (phononic crystal) matrix\cite{liu2000locally}. If the inclusion is properly chosen so that it presents low frequency resonances, these structures behave like effective materials with resonant-like constitutive parameters which can be either positive, zero or negative \cite{li2004double}.

Phononic and sonic crystals are anisotropic structures in general, therefore they present anisotropic constitutive parameters. Then, it was demonstrated that metafluids present anisotropic mass density not only near a local resonance, but also in the quasi-static limit\cite{torrent2008anisotropic,torrent2011multiple}, although for the case of metasolids it has been assumed in general that the mass density is a scalar\cite{wu2007effective,zhou2009analytic,lai2011hybrid}. Recently, some works have shown that elastic composites have to be described by means of the so-called ``Willis form'' of the constitutive parameters\cite{milton2006cloaking,milton2007modifications}, which include a tensorial mass density and an additional coupling tensor, and the dynamic homogenization of phononic crystals has also shown that this general description applies to these structures\cite{norris2012analytical}.

In this work the low-frequency limit of phononic crystals is analyzed, and it is shown analytically that low-frequency resonances actually induce an anisotropic mass density, which however becomes scalar in the static limit. It is also shown that the usual assumption that the negative mass density is induced by dipolar resonances is not necessarily true, in a similar way as was previously demonstrated for metamaterials for flexural waves by the authors in a recent publication\cite{torrent2014effective}. Finally, the dispersion relation of the full phononic crystal is compared with that of a homogeneous material with the obtained effective parameters, and a good agreement is found in general, although it is also demonstrated that non-local parameters have to be considered.

The paper is organized as follows: After this introduction, Section \ref{sec:homo} explains the homogenization method employed here. Following, Section \ref{sec:perturbation} explains how to apply perturbation theory to derive some important properties of metasolids in the low frequency limit. Finally, Section \ref{sec:local} describes a phononic crystal as a locally resonant metamaterial and Section \ref{sec:numerical} shows a numerical example of application of the theory. Section \ref{sec:summary} summarizes the work.

\section{Homogenization of the Periodic Medium from the Band Structure}
\label{sec:homo}
The equation of motion of an inhomogeneous solid, assuming harmonic time dependence with frequency $\omega$, is given by the classical elastodynamic equation \cite{royer2000elastic}
\beq
-\rho(\fr) \omega^2 u_i=\partial_j C_{ijkl}(\fr)\partial_k u_l
\eeq
being $u_i$ the components of the displacement field and $C_{ijkl}$ the components of the stiffness tensor. Hereafter we employ the summation convention in which two repeated indexes implies summation over their possible values, to simplify notation. If he medium is homogeneous the dispersion relation is obtained by assuming plane-wave propagation with wavevector $\bm{k}=k\bm{n}$, with $k$ being the wavenumber and $\bm{n}$ a unit vector parallel to de propagation direction. Under this assumption, the equation of motion becomes the well-known secular equation for elastic waves \cite{royer2000elastic}
\beq
\label{eq:homo}
\rho \omega^2 u_i=k^2n_{iI} C_{IJ}n_{Jj} u_j,
\eeq
with $u_i$ being the components of the displacement field and $C_{IJ}$ the components of the stiffness tensor in Voigt notation (see Ref. \onlinecite{royer2000elastic} and the Appendix \ref{sec:notation}). The solution for the dispersion relation $\omega=\omega(k,\bm{n})$ is therefore given by the roots of the determinant of the matrix $\bar{\Gamma}$ defined as
\beq
\bar{\Gamma}_{ij}=\rho \omega^2\delta_{ij}-k^2n_{iI} C_{IJ}n_{Jj}.
\eeq

In a phononic crystal both $\rho(\fr)$ and $C_{ijkl}(\fr)$ are periodic functions of the spatial coordinates, then Bloch theorem is applied and the Plane Wave Expansion method\cite{kushwaha1993acoustic} can be used to obtain the dispersion relation $\omega=\omega(k,\bm{n})$ as the solution of the following eigenvalue equation
\beq
\omega^2\rho_{\fG-\fG'}(u_{\fG'})_i=(k+G)_{iI}C_{IJ}^{\fG-\fG'}(k+G')_{Jj}(u_{\fG'})_j
\eeq
where $\rho_{\fG},C_{IJ}^{\fG}$ and $(u_{\fG})_i$ stands for the Fourier components of the mass density, stiffness tensor and displacement field, respectively and summation over repeated indexes is assumed including the Fourier indexes defined by the reciprocal lattice vector $\fG$. The matrix elements $(k+G)_{iI}$ are defined in Appendix \ref{sec:notation}. In matrix form the above eigenvalue equation is expressed as
\beq
\label{eq:NGMG}
\omega^2 N_{\fG\fG'}\bm{u}_{\fG'}=M_{\fG\fG'}\bm{u}_{\fG'}.
\eeq
where
\begin{align}
(N_{\fG\fG'})_{ij}&=\rho_{\fG-\fG'}\delta_{ij} \label{eq:NGGp},\\
(M_{\fG\fG'})_{ij}&=(k+G)_{iI} C^{\fG-\fG'}_{IJ}(k+G')_{Jj}\label{eq:MGGp}.
\end{align}

This equation solves for the dispersion relation inside the phononic crystal, however in its current form it is difficult to figure out any property of the  crystal as a composite. The description of the crystal as a material can be obtained by averaging the components of the displacement vector in the unit cell and finding in this way an equation similar to equation \eqref{eq:homo}, in which the coefficients of the different terms multiplying the wavevector and the frequency define the effective parameters. The average of the displacement vector is given by the $\fG=0$ component of $\fu_{\fG}$, so that it can be obtained by expressing equation \eqref{eq:NGMG} as
\begin{subequations}
\label{eq:eigexpanded}
\begin{align}
\omega^2 N_{00}\fu_0+\omega^2 N_{0\fG'}\fu_{\fG'}&=M_{00}\fu_0+M_{0\fG'}\fu_{\fG'}\\
\omega^2 N_{\fG 0}\fu_0+\omega^2 N_{\fG\fG'}\fu_{\fG'}&=M_{\fG 0}\fu_0+M_{\fG\fG'}\fu_{\fG'}.
\end{align}
\end{subequations}
It must be recalled that repeated indexes means summation over all their possible values, and that hereafter it is considered that matrix elements labelled with $\fG$ does not include the term $\fG=0$, which is extracted from the above decomposition. We can now solve from the second equation for $\fu_{\fG}$,
\beq
\fu_{\fG'}=-(M_{\fG' \fG}-\omega^2 N_{\fG' \fG})^{-1}(M_{\fG 0}-\omega^2 N_{\fG 0})\fu_0
\eeq
and insert it into the first one, obtaining the following equation
\begin{multline}
\label{eq:secext}
\left[\omega^2 N_{00}-\omega^2 N_{0\fG'}\chi_{\fG'\fG}(M_{\fG 0}-\omega^2 N_{\fG 0})\right.\\
\left. -M_{00}+M_{0\fG'}\chi_{\fG'\fG}(M_{\fG 0}-\omega^2 N_{\fG 0})\right]\fu_0=0
\end{multline}
where we have defined
\beq
\label{eq:chiomegak}
\chi_{\ell m}^{\fG'\fG}(\omega,\bm{k})\equiv (M_{\fG' \fG}-\omega^2 N_{\fG' \fG})^{-1}_{\ell m}.
\eeq

Equation \eqref{eq:secext} is formally the same as equation \eqref{eq:NGMG}, however it is not an eigenvalue equation, but a secular equation for $\fu_0$ similar to equation \eqref{eq:homo}, where the solutions $\omega=\omega(k,\bm{n})$ are obtained from the zeros of the determinant of the matrix $\Gamma$ defined as
\begin{multline}
\Gamma=\omega^2 N_{00}-\omega^2 N_{0\fG'}\chi_{\fG'\fG}(M_{\fG 0}-\omega^2 N_{\fG 0})\\
-M_{00}+M_{0\fG'}\chi_{\fG'\fG}(M_{\fG 0}-\omega^2 N_{\fG 0}).
\end{multline}

The matrix $\Gamma$ is actually a $3\times 3$ matrix, in which coefficients are in general functions of both $\omega$ and $\bm{k}$, what makes it less suitable for band structure calculation than equation \eqref{eq:NGMG} but more suitable for the description of the phononic crystal as a composite. Effectively, we can see that the elements of the $N_{00}, N_{0\fG'}$ and $N_{\fG0}$ does not depend explicitly on the wavevector $\bm{k}$, 
\begin{align}
(N_{00})_{ij}&=\bar{\rho}\delta_{ij},\\
(N_{0\fG'})_{ij}&=\rho_{-\fG'}\delta_{ij},\\
(N_{\fG0)_{ij}}&=\rho_{\fG}\delta_{ij},
\end{align}
while the $M_{00}, M_{0\fG'}$ and $M_{\fG0}$ contains this dependence,
\begin{align}
(M_{00})_{ij}&=k_{iI} \bar{C}_{IJ}k_{Jj},\\
(M_{0\fG'})_{ij}&=k_{iI} C^{\fG-\fG'}_{IJ}(k+G')_{Jj},\\
(M_{\fG0)_{ij}}&=(k+G)_{iI} C^{\fG-\fG'}_{IJ}k_{Jj}.
\end{align}
The dependence with the wavevector and frequency can be reorganized then the $\Gamma$ matrix can be cast as
\beq
\label{eq:Mnonlocal}
\Gamma_{ij}=\omega^2\rho_{ij}^*-k^2n_{iI}C_{IJ}^*n_{Jj}-\omega k(n_{iI}S_{Ij}+S_{iJ}^\dag n_{Jj})
\eeq
where the coefficients $\rho_{ij}^*, C_{IJ}^*$ and $S_{Ij}$ are given by
\begin{subequations}
\label{eq:rhoCSNL}
\begin{align}
\rho_{ij}^{*}(\omega,\bm{k})&=\bar{\rho}\delta_{ij}+\omega^2\rho_{-\fG'}\chi_{ij}^{\fG'\fG}(\omega,\bm{k})\rho_{\fG}\\
C_{IJ}^*(\omega,\bm{k})&=\bar{C}_{IJ}-\nonumber\\
&C_{IL}^{-\fG'}(k+G')_{L\ell}\chi_{\ell m}^{\fG'\fG}(\omega,\bm{k})(k+G)_{mM}C_{MJ}^{\fG}\\
S_{Ij}	(\omega,\bm{k})&=\omega C_{IL}^{-\fG'}(k+G')_{L\ell}\chi_{\ell j}^{\fG'\fG}(\omega,\bm{k})\rho_{\fG}
\end{align}
\end{subequations}

Equation \eqref{eq:Mnonlocal} is similar to equation \eqref{eq:homo}, but the constitutive parameters required to describe the phononic solid are more complex. This equation shows that the phononic crystal is a non-local Willis medium \cite{milton2007modifications,norris2012analytical}, in which the mass density is a tensorial quantity and with the presence of the coupling field $S_{Ij}$. The above expressions are valid at any frequency and wavenumber, however in this work we are specially interested in the low frequency limit, that is, the limit in which the wavelength of the field in the background is larger than the typical periodicity of the crystal and it is described as a homogeneous material. It will be shown that even in the low-frequency limit these systems can have some special resonances in which the crystal behaves like a Willis medium with resonant and non-local parameters.

If in equations \eqref{eq:rhoCSNL} the limit $\omega\to0$ and $k\to 0$ is taken it is found that the coupling field $S_{Ij}=0$, since it is directly proportional to $\omega$. Also, the mass density becomes a scalar which is simply the volume average $\rho_{ij}=\bar{\rho}\delta_{ij}$, as it is well known from the theory of composites. Finally, the effective stiffness tensor is given by
\beq
C_{IJ}^*=\bar{C}_{IJ}-C_{IL}^{-\fG'}G'_{L\ell}(M^{-1}_{\fG'\fG})_{\ell m}G_{mM}C_{MJ}^{\fG}
\eeq
and the medium behaves like an effective homogeneous medium with local and frequency-independent parameters. The mass density is a scalar, therefore all the information about the microstructure of the composite is contained in the $C_{IJ}^*$ tensor, whose symmetry will depend on the background, inclusions and lattice symmetry.
In the above expression it has been assumed the limit in which the frequency and the wavenumber tend to zero, however in practice this limit will be valid from  zero to some cut-off frequency in which it will not be possible to neglect some terms containing the frequency or the wavenumber, the medium begins then to be dispersive and the constitutive parameters will depend on both frequency and wavenumber. 
It can happen however that these parameters be frequency-dependent even in the low frequency limit, under the condition what is called a ``local resonance''. This happens when the parameter $\chi_{\fG\fG'}$ is singular, then it is found that all the constitutive parameters can become resonant and locally singular, 
 with the remarkable result that it presents, depending on the lattice symmetry, anisotropic mass density.
\section{Perturbation Theory in the Low Frequency Limit}
\label{sec:perturbation}
 The  origin of the resonant parameters of these structures is the $\chi$ matrix defined in general as
\beq
\chi=(M-\omega^2 N)^{-1}.
\eeq
It must be pointed out that, in the static limit this $\chi$ matrix simply is the reciprocal of the matrix $M$, however the term $\omega^2 N$ can make that the determinant of the matrix $M-\omega^2 N$ be zero for some specific values of $\omega$, which we call resonances because then the $\chi$ matrix is singular. Let us try to understand the nature of these resonances.

Let us assume that we know the eigenvalues $\lambda_n$ and eigenvectors $\fv_n$ of the matrix $M-\omega^2 N$. We know then that the reciprocal of this matrix can be expanded by means of the eigen-decomposition theorem, thus we have that
\beq
\chi=(M-\omega^2 N)^{-1}=\sum_n \frac{\fv_n^\dag \otimes \fv_n}{\lambda_n}
\eeq
given that $M-\omega^2N$ is actually a Hermitian matrix. The matrix $M$ is defined in equation \eqref{eq:MGGp}, and it can be expressed as
\beq
M=M_0+kM_1+k^2M_2,
\eeq
where
\begin{align}
M_0&=G_{iI} C^{\fG-\fG'}_{IJ}G'_{Jj},\\
M_1&=n_{iI} C^{\fG-\fG'}_{IJ}G'_{Jj}+G_{iI} C^{\fG-\fG'}_{IJ}n_{Jj},\\
M_2&=n_{iI} C^{\fG-\fG'}_{IJ}n_{Jj}.
\end{align}
For low frequencies and wavenumbers, the matrix $M$ can be considered a perturbation of the $M_0$ matrix, so that we can apply perturbation theory to relate the eigenvalues $\lambda_k$ with frequency. Let us define $\fu_n$ and $C_n^0/a^2$ (with $a$ being a quantity with units of length, for convenience in the units) the eigenvectors and eigenvalues of the $M_0$ matrix, respectively, thus
\beq
M_0\fu_n=C_n^0/a^2\fu_n,
\eeq
if we assume that $kM_1+k^2M_2-\omega^2N$ is a perturbation of the matrix $M_0$, the eigenvalues and eigenvectors $\lambda_n$ and $\fv_n$ will be given, up to first order in perturbation theory, by
\begin{align}
\lambda_n&=C_n^0/a^2+k C_n^{(1)}/a+k^2 C_n^{(2)}-\omega^2\rho_n,\\
\fv_n&=\fu_n+k\sum_{\ell}b_{n\ell}^{(1)}\fu_\ell+k^2\sum_{\ell}b_{n\ell}^{(2)}\fu_\ell-\omega^2\sum_{\ell}a_{n\ell}\fu_\ell,
\end{align}
where (assuming $\fu_n\cdot\fu_n=1$)
\begin{align}
\rho_n&=\fu_n^\dag N\fu_n,\\
C_n^{(1)}/a&=\fu_n^\dag M_1\fu_n,\\
C_n^{(2)}&=\fu_n^\dag M_2\fu_n,
\end{align}
and, for $n\neq \ell$,
\begin{align}
a_{n\ell}&=\frac{\fu_n^\dag N\fu_\ell}{C_n^0/a^2-C_\ell^0/a^2},\\
b_{n\ell}^{(1)}&=\frac{\fu_n^\dag M_1\fu_\ell}{C_n^0/a^2-C_\ell^0/a^2},\\
b_{n\ell}^{(2)}&=\frac{\fu_n^\dag M_2\fu_\ell}{C_n^0/a^2-C_\ell^0/a^2},
\end{align}
which ensures as well that $\fv_n\cdot\fv_n=1$. 

The dependence in $k$ of the eigenvalues $\lambda_n$ implies that the effective parameters will be non-local in general, however, as it will be shown later, metamaterials are in general designed by means of ``soft'' scatterers, that is, it is required that the velocity of the waves inside the scatterers be much smaller than that of the background, in other words, the components of the stiffness matrix are in general much smaller than the density. Therefore, as a first approximation, we can neglect the coefficients multiplying the wavenumber and approximate $\lambda_n$ as
\beq
\lambda_n\approx C_n^0/a^2-\omega^2\rho_n,\\
\eeq
and allows expressing the $\chi$ matrix as (neglecting the perturbative terms in $\fu$)
\beq
\label{eq:chiexpanded}
\chi_{ij}^{\fG'\fG}(\omega)=\sum_n \frac{(u_n^*)_{\fG' i}  (u_n)_{\fG j}}{C_n^0/a^2-\omega^2\rho_n}.
\eeq
This interesting result shows that at the resonant frequencies $\omega_n^2 a^2=C_n/\rho_n$ the effective parameters become singular, and in the neighborhood of this frequency the can have positive, negative or zero values. These resonances are determined by the ratio of the eigenvalues $C_n$ of the matrix $M_0$ and by the perturbation term $\rho_n$. Also, the coupling of these resonances with the different constitutive parameters is defined by the symmetry of the eigenvectors $\fu_n$ of the matrix $M_0$, as will be explained in the following section. 

It must be mentioned that the eigenvectors $\fu_n$ correspond to the eigenvectors of the matrix $M_0$, which actually is the matrix $M_{\fG\fG'}$ given by equation \eqref{eq:MGGp} but for $k=0$ and removing the terms corresponding to $\fG=0$. These eigenvectors correspond to a physical system which is easy to understand: Imagine a phononic crystal in which the stiffness tensor is periodic while the mass density is equal to that of the background. It is easy to see that the eigenvalue equation of this system at the $\Gamma$ point, that is, for $k=0$, is according to equations \eqref{eq:eigexpanded}, 
\begin{subequations}
\begin{align}
\omega^2 \fu_0&=0\\
\omega^2\fu_{\fG'}&=M_{\fG\fG'}\fu_{\fG'}.
\end{align}
\end{subequations}
which for $\omega\neq 0$ has the only solution $\fu_{\fG}=(0,\fu_n)$, being $\fu_n$ the eigenvectors of the matrix $M_0$. This relationship between the eigenvalues and eigenvectors of a physical system with those required to compute the effective parameters suggest that other numerical methods more powerful than the Plane Wave Expansion method could be used to characterize these systems, after properly Fourier transform solutions and elastic constants distribution. Therefore, on the basis of the present theory, this work opens a door to a more efficient calculation of the effective parameters, whose discussion is beyond the objective of the present work.

%
\section{Effective Parameters in the Local Approximation}
\label{sec:local}

The study of a local resonance is made in a regime in which the wavelength of the propagating field is larger than the typical periodicity of the composite. This hypothesis implies that in equations \eqref{eq:rhoCSNL} we can make the approximation $k+G\approx G$, so that in these equations the dependence on $k$ disappears and the parameters, although frequency-dependent, are ``local'' and given by
\begin{subequations}
\label{eq:rhoCSL}
\begin{align}
\rho_{ij}^{*}(\omega)&=\bar{\rho}\delta_{ij}+\omega^2\rho_{-\fG'} \chi_{\ell m}^{\fG'\fG}(\omega) \rho_{\fG}\\
C_{IJ}^*(\omega)&=\bar{C}_{IJ}-C_{IL}^{-\fG'}G'_{L\ell}\chi_{\ell m}^{\fG'\fG}(\omega)G_{mM}C_{MJ}^{\fG}\\
S_{Ij}(\omega)&=\omega C_{IL}^{-\fG'}G'_{L\ell}\chi_{\ell j}^{\fG'\fG}(\omega)\rho_{\fG}
\end{align}
\end{subequations}
where $\chi_{ij}^{\fG'\fG}(\omega)$ is computed from equation \eqref{eq:chiomegak} as $\chi_{ij}^{\fG'\fG}(\omega,\bm{k}=0)$. A better insight in the properties of these parameters can be done by including the expansion of $\chi$ given by equation \eqref{eq:chiexpanded}, then we have
\begin{subequations}
\label{eq:rhoCSLbis}
\begin{align}
\rho_{ij}^{*}(\omega)&=\bar{\rho}\delta_{ij}+\sum_n \omega^2\rho_{-\fG'} \frac{(u_n^*)_{\fG' i}  (u_n)_{\fG j}}{C_n^0/a^2-\omega^2\rho_n} \rho_{\fG},\\
C_{IJ}^*(\omega)&=\bar{C}_{IJ}-\nonumber \\
&\sum_n C_{IL}^{-\fG'}G'_{L\ell}\frac{(u_n^*)_{\fG' \ell}  (u_n)_{\fG m}}{C_n^0/a^2-\omega^2\rho_n}G_{mM}C_{MJ}^{\fG},\\
S_{Ij}(\omega)&=\sum_n\omega C_{IL}^{-\fG'}G'_{L\ell}\frac{(u_n^*)_{\fG' \ell}  (u_n)_{\fG j}}{C_n^0/a^2-\omega^2\rho_n}\rho_{\fG}.
\end{align}
\end{subequations}
We can now define the quantities
\begin{align}
(A_n)_i&= (u_n)_{\fG i} \rho_{\fG}, \\
(B_n)_I&=(u_n)_{\fG m} G_{mM} C_{MI}^{\fG},
\end{align}
and, given that any Fourier coefficient satisfies $F_{-\fG}=F_{\fG}^*$ we get for the effective parameters  the following expressions
\begin{subequations}
\label{eq:rhoCSLfinal}
\begin{align}
\rho_{ij}^{*}(\omega)&=\bar{\rho}\delta_{ij}+\omega^2\sum_n\frac{(A_n^*)_i(A_n)_j}{C_n^0/a^2-\omega^2\rho_n},\label{eq:rhoused}\\
C_{IJ}^*(\omega)&=\bar{C}_{IJ}-\sum_n\frac{(B_n^*)_I(B_n)_J}{C_n^0/a^2-\omega^2\rho_n},\\
S_{Ij}(\omega)&=\omega \sum_n\frac{(B_n^*)_I(A_n)_j}{C_n^0/a^2-\omega^2\rho_n}.
\end{align}
\end{subequations}
The above equations relate the effective parameters with the properties of the eigenvectors and eigenvalues of the matrix $M_0$, as well as with their perturbations. Let us note that, for a symmetric unit cell, we will have that $F(-\fG)=F(\fG)$, and it is easy to understand that for this type of systems we have that
\beq
\fu_{-\fG}=\pm\fu_{\fG},
\eeq
which implies two type of solutions, $(A_n)_{i}\neq 0$ and $(B_n)_{I}=0$ when $\fu_{-\fG}=\fu_{\fG}$ and $(A_n)_{i}=0$ and $(B_I)_{n}\neq 0$ when $\fu_{-\fG}=-\fu_{\fG}$. The former induces a resonant mass density, while the latter induces a resonant stiffness tensor. The two cases implies that the Willis tensor $S_{Ij}$ is equal to zero, and this also implies that a symmetric system cannot excite simultaneously a resonance in the stiffness tensor and the mass density, that is, we cannot have double negative materials in this way, unless the different resonances be too close each other.

A deeper insight into the properties of symmetry and non-symmetry of these resonances is beyond the objective of the present work, in which we want to focus attention on the properties of the resonant mass density, however a future work concerning non-symmetric lattice will be prepared and published elsewhere.


\section{Numerical Example: Resonant and Non-Local Anisotropic Mass Density}
\label{sec:numerical}

\begin{figure}
\centering
\includegraphics[width=0.5\columnwidth]{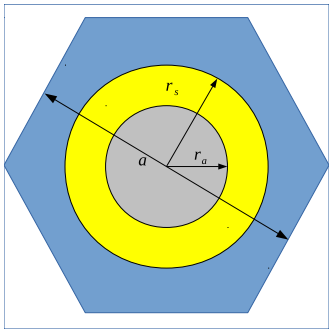}
\caption{\label{Fig:uniccell} Phononic crystal studied in the present work. The system consist of a triangular arrangement of coated cylinders in an epoxy background. The cylinders consist of a lead core of radius $r_a$ and a rubber shell of radius $r_s$ (see text for numerical values).}
\end{figure}

Figure \ref{Fig:uniccell} shows the system to be studied in the present work. It consists in a periodic arrangement of coated cylinders in an epoxy background ($\rho_b=1.18$ Kg/dm$^3$, $E_b=4.35$ GPa and $\nu_b=0.37$). The cylinders are made of a lead core ($\rho_a=11.34$ Kg/dm$^3$, $E_a=16$ GPa and $\nu_a=0.44$) of radius $r_a=0.16a$ and a rubber shell ($\rho_s=1.3$ Kg/dm$^3$, $E_s=2.7E-4$ GPa and $\nu_b=0.499$) of radius $r_s=0.4a$, since this combination of soft-hard coatings is known to present low frequency resonances. The cylinders are arranged in a triangular lattice, in this way we expect the effective material to be transversely isotropic.


\begin{figure}
\centering
\includegraphics[scale=1]{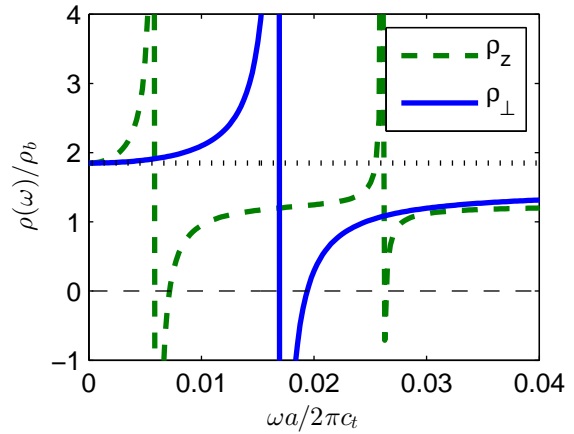}
\caption{\label{fig:rhoL} Effective mass density tensor for the proposed phononic crystal. Both the transversal component (blue continuous line) and the z component (green dashed line) present a local resonance. In the vicinity of this resonance both components of the mass density become negative. }
\end{figure}

Figure \ref{fig:rhoL} shows the effective mass density tensor relative to that of the background $\rho_b$ as a function of frequency as computed by using equation \eqref{eq:rhoused}. The symmetry of the lattice makes that any second rank tensor will have only two components, one for the $xy$ plane and another one for the $z$ plane. It is seen how in the low frequency limit the two components are identical and equal to the normalized average mass density $\bar{\rho}/\rho_b$, as expected, however it can also be seen how they split as a function of frequency and present two different resonances, so that the system behaves like an elastic medium with anisotropic mass density. Moreover, it can also be seen how these components are negative in different frequency regions. It is found that the effective stiffness tensor is nearly constant in frequency in this region, and that the coupling field $S_{Ij}$ is zero, as expected from the discussion in the previous section.

\begin{figure}
\centering
\includegraphics[width=0.7\columnwidth]{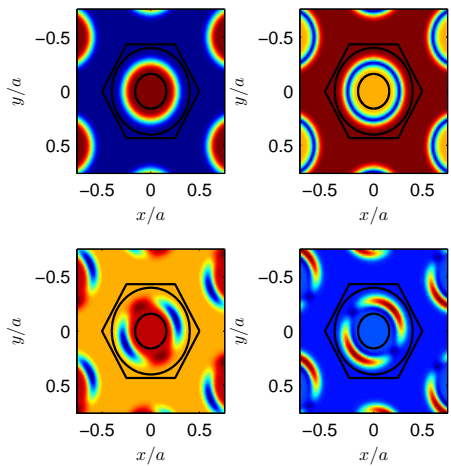}
\caption{\label{Fig:modes} Resonant modes inducing an effective negative mass density. Upper panels for $\omega_k a/2\pi c_t=0.0072$, corresponding to a resonance in $\rho_z$, lower panels for $\omega_k a/2\pi c_t=0.0194$, corresponding to a resonance in $\rho_\perp$ (see text for further discussion).}
\end{figure}

Figure \ref{Fig:modes} shows the field distribution of the lower frequency resonances of the mass density tensor depicted in figure \ref{fig:rhoL}, the upper panel for the z component and the lower panel for the xy one, left panels show the real part of the mode while right panels show the absolute value. It is interesting to note that the xy mode has a dipolar symmetry, as it is commonly assumed in the literature \cite{zhou2009analytic}, while the z mode has monopolar symmetry. The fact that a monopolar symmetry could induce a negative mass density behaviour was already found by the authors in a recent paper \cite{torrent2014effective} in the study of flexural waves in thin plates. This result is consistent with the theory of elastic waves in plates, given that a plate with a periodic arrangement of inclusions is indeed a finite slide of the two-dimensional phononic crystal studied here, and this result suggest that the propagation of flexural waves is mainly dominated by the z component of the mass density. This important result should be taken into account in the homogenization theory of plate metamaterials, although a deep insight into it is beyond the objective of the present work.

Equations \eqref{eq:rhoCSL} show then that the phononic crystal can be described by means of locally resonant constitutive parameters, whose frequency dependence can be easily computed. The description of a phononic crystal as a frequency-dependent homogeneous material will not be valid for every wavenumber and frequency, and to determine these limits the dispersion relation obtained by means of the constitutive parameters is compared with the band structure obtained from the eigenvalue equation \eqref{eq:NGMG}. 
Given that $S_{Ij}$ is zero for this example and $C_{IJ}^*$ is constant in frequency, along the $\Gamma X$ direction the dispersion relation for the effective material is
\begin{align}
\omega^2\rho_\perp(\omega) u_x&=k_x^2C_{11}^*u_x,\\
\omega^2\rho_\perp(\omega) u_y&=k_x^2C_{66}^*u_y,\\
\omega^2\rho_z(\omega) u_z&=k_x^2C_{44}^*u_z,
\end{align}
while along the $\Gamma A$ direction, that is, along the z axis, the dispersion relation is (notice that in this case $C_{55}^*=C_{44}^*$)
\begin{align}
\omega^2\rho_\perp(\omega) u_x&=k_z^2C_{44}^*u_x,\\
\omega^2\rho_\perp(\omega) u_y&=k_z^2C_{55}^*u_y,\\
\omega^2\rho_z(\omega) u_z&=k_z^2C_{33}^*u_z.
\end{align}

\begin{figure}
\centering
\includegraphics[width=0.9\columnwidth,height=6cm]{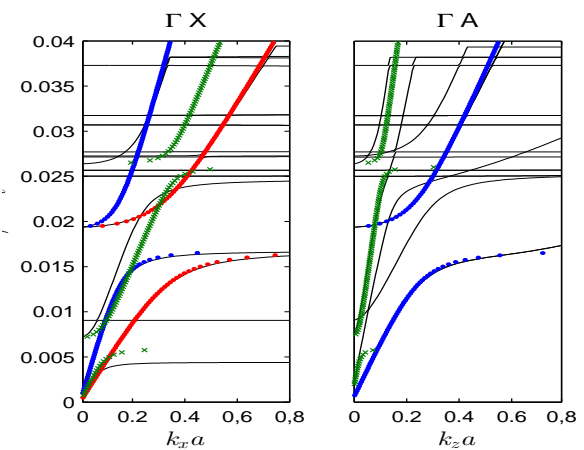}
\caption{\label{Fig:GXGA} Left panel: Dispersion relation of the phononic crystal along the $\Gamma X$ direction (black lines) compared with those obtained from the effective local constitutive parameters (green crosses and blue and red dots). Right panel: Dispersion relation of the phononic crystal along the $\Gamma A$ direction (black lines) compared with those obtained from the effective local constitutive parameters (green crosses and blue dots).}
\end{figure}

Figure \ref{Fig:GXGA}, left panel, shows the dispersion relation along the $\Gamma X$ direction (x axis) computed by means of the eigenvalue equation \eqref{eq:NGMG} (black lines) compared with the dispersion relation obtained by means of the constitutive parameters. Red and blue dots show the results for the xy modes, and it is seen that there is a good agreement between the eigenvalue equation and the effective material dispersion relation. The dispersion relation for the z mode (green crosses) is however different from the eigenvalue equation and the effective material, and there is an agreement only for very low wavenumbers. As will be seen later, the reason for this disagreement is that the local description of the metamaterial is not accurate here, and it is required the inclusion of the non-local components, that is, the dependence on the wavenumber in the constitutive parameters.

Figure \ref{Fig:GXGA}, right panel, shows similar results for propagation along the $\Gamma A$ direction (z axis). It is shown here that the xy modes, which are degenerate given that the crystal is transversely isotropic, are perfectly described by means of the effective material, however the z modes, corresponding to green crosses, agree only for very low wavenumbers. There are also a set of flat bands that can be fairly difficult to predict by means of the effective material parameters. The reason for that is that these modes occur only at a given frequency and correspond to very sharp modes, and although they are properly predicted by the theory as a resonant frequency $\omega_n$, their effect is difficult to see in the constitutive parameters.

\begin{figure}
\centering
\includegraphics[width=0.8\columnwidth, height=8cm]{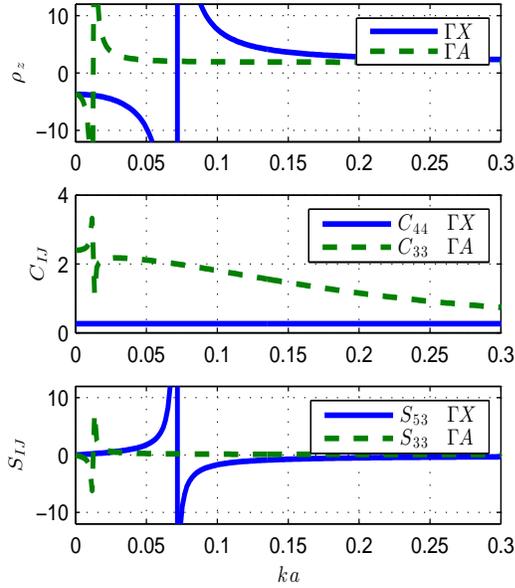}
\caption{\label{Fig:NL} Non-local constitutive parameters related with the propagation of the z mode in the phononic crystal. Upper panel: z component of the mass density. Mid panel: $C_{44}$ and $C_{33}$ components of the stiffness tensor. Lowe panel: $S_{53}$ and $S_{33}$ components of the coupling field (see text for details).}
\end{figure}

The spatial dispersion of the z mode can be understood by means of the calculation of the non-local constitutive parameters using equations \eqref{eq:rhoCSNL}. Figure \ref{Fig:NL} shows these parameters at a frequency $\omega a/2\pi c_t=0.0028$, corresponding to a frequency in which the z component of the local mass density is negative. The upper panel shows the non-local $\rho_z$ as a function of the wavenumber along the $\Gamma X$ and $\Gamma A$ directions. It is clear that the origin of the non-locality is a wavenumber resonance, for which the major contribution will have its origin in the $\chi$ matrix. It is also seen that the $C_{33}$ component, responsible of the propagation of the mode along the $\Gamma A$ direction, also becomes non-local, while the $C_{44}$ remains constant. Additionally, the $S_{53}$ and $S_{33}$ elements, which are zero for $k=0$, appear as resonant components. The contribution of these spatial resonances is essentially to displace the opening of the band gaps, as can be seen from figure \ref{Fig:GXGA}, for which their influence is important before considering only the local theory. 
\section{Summary}
\label{sec:summary}

In summary, it has been analytical and numerically demonstrated that phononic crystals behave as elastic metasolids with anisotropic, resonant and non-local effective parameters, with the remarkable result that the mass density is also anisotropic in general, although in the static limit this quantity recovers its scalar nature. Also, it has been demonstrated that the symmetry of the resonance inducing this behaviour is not necessarily dipolar, as it is commonly assumed, while it can also be monopolar. The non-local and anisotropic nature of the mass density has important implications specially for the study of plate metamaterials, since these structures are essentially finite slides of phononic crystals. It must be pointed out that the generality of the equations derived can be used to the homogenization of phononic crystals with more complex unit cells, with the objective of achieving double negative metasolids. Finally, the theory can be extended to phononic crystals with piezoelectric inclusions, where resonant piezoelectric constants are expected.
\section*{ACKNOWLEDGMENT}
This work was supported by the ``Agence Nationale de la Recherche (ANR)'' and the ``D\'el\'egation G\'en\'erale a l'Armement (DGA)'' under the project Metactif, Grant No. ANR-11-ASTR-015 and by the LabEx AMADEus (ANR-10-LABX-42) in the framework of IdEx Bordeaux (ANR-10-IDEX-03-02), France.

\appendix
\section{Matrix Notation}
\label{sec:notation}
Through the paper Voigt notation for the indexes is used, in such a way that lower-case indexes run from 1 to 3 and upper case indexes run from 1 to 6. Also, the wavevector is defined in terms of the $V_{iI}$ matrix defined as
\beq
\bm{V}=\left(
\begin{matrix}
V_x & 0   & 0   & 0   & V_z & V_y\\
0   & V_y & 0   & V_z & 0   & V_x\\
0   & 0   & V_z & V_y & V_x & 0
\end{matrix}
\right)
\eeq
Therefore the matrix elements $(k+G)_{iI}$ are
\begin{widetext}
\beq
\bm{k+G}=\left(
\begin{matrix}
k_x+G_x & 0   & 0   & 0   & k_z+G_z & k_y+G_y\\
0   & k_y+G_y & 0   & k_z+G_z & 0   & k_x+G_x\\
0   & 0   & k_z+G_z & k_y+G_y & k_x+G_x & 0
\end{matrix}
\right)
\eeq
\end{widetext}
being therefore $(k+G)_{Jj}$ the transpose of the above matrix. Similarly, the same matrix $n_{iI}$ is defined for the normal vector $\bm{n}$, 
\beq
\bm{n}=\left(
\begin{matrix}
n_x & 0   & 0   & 0   & n_z & n_y\\
0   & n_y & 0   & n_z & 0   & n_x\\
0   & 0   & n_z & n_y & n_x & 0
\end{matrix}
\right)
\eeq

\end{document}